\pdfoutput=1

\documentclass[11pt,twoside,a4paper,cmspaper,final,collab]{cms-tdr}
\def\svnVersion{fab8662}\def\svnDate{2019/08/22}\def\cmsCernNoTag{CERN-EP-2019-113}\def\cmsCernDate{\today}\def\cmsMessage{"Published in Physics Letters B as \href{http://dx.doi.org/10.1016/j.physletb.2019.134876}{\doi{10.1016/j.physletb.2019.134876}.}"}

\begin{document}\cmsNoteHeader{EXO-19-001}

\hyphenation{had-ron-i-za-tion}
\hyphenation{cal-or-i-me-ter}
\hyphenation{de-vices}
\RCS$HeadURL: svn+ssh://svn.cern.ch/reps/tdr2/notes/AN-18-332/trunk/AN-18-332.tex $
\RCS$Id: AN-18-332.tex 339081 2016-04-18 15:39:58Z sakuma $

\newcommand{\ctau}{\ensuremath{c\tau_{0}}\xspace}
\newcommand{\ate}{\ensuremath{\mathcal{A}\varepsilon}\xspace}
\newcommand{\emj}{\ensuremath{E_{\textrm{ECAL}}}\xspace}
\newcommand{\nec}{\ensuremath{N^{\textrm{cell}}_{\textrm{ECAL}}}\xspace}
\newcommand{\hmj}{\ensuremath{E_{\textrm{HCAL}}}\xspace}
\newcommand{\jt}{\ensuremath{t_{\textrm{jet}}}\xspace}
\newcommand{\jtrms}{\ensuremath{t^{\textrm{RMS}}_\textrm{jet}}\xspace}
\newcommand{\hef}{\ensuremath{\textrm{HEF}}\xspace}
\newcommand{\cscef}{\ensuremath{E^{\textrm{CSC}}_\textrm{ECAL}/E_{\textrm{ECAL}}}\xspace}
\newcommand{\dtmaxdphi}{\ensuremath{\max(\Delta \phi_{\mathrm{DT}})}\xspace}
\newcommand{\rpcdtmaxdphi}{\ensuremath{\max(\Delta \phi_{\mathrm{DT/RPC}})}\xspace}
\newcommand{\rpcmaxdphi}{\ensuremath{\max(\Delta \phi_{\mathrm{RPC}})}\xspace}
\newcommand{\jtrmsfrac}{\ensuremath{\jtrms/\jt}\xspace}
\newcommand{\pttf}{\ensuremath{{\mathrm {PV}}_{\text{track}}^{\text{fraction}}}\xspace}
\newcommand{\mg}{\ensuremath{m_{\PSg}}\xspace}
\newlength\cmsFigWidth
\ifthenelse{\boolean{cms@external}}{\setlength\cmsFigWidth{\columnwidth}}{\setlength\cmsFigWidth{0.7\textwidth}}
\ifthenelse{\boolean{cms@external}}{\providecommand{\cmsLeft}{upper figure\xspace}}{\providecommand{\cmsLeft}{left figure\xspace}}
\ifthenelse{\boolean{cms@external}}{\providecommand{\cmsRight}{lower figure\xspace}}{\providecommand{\cmsRight}{right figure\xspace}}
{\providecommand{\CL}{CL\xspace}

\cmsNoteHeader{EXO-19-001}

\title{Search for long-lived particles using nonprompt jets and missing transverse momentum with proton-proton collisions at \texorpdfstring{$\sqrt{s}=13\TeV$}{sqrt(s)=13 TeV}}

\date{\today}

\abstract{A search for long-lived particles decaying to displaced, nonprompt jets and missing transverse momentum is presented. The data sample corresponds to an integrated luminosity of 137\fbinv of proton-proton collisions at a center-of-mass energy of 13\TeV collected by the CMS experiment at the CERN LHC in 2016--2018. Candidate signal events containing nonprompt jets are identified using the timing capabilities of the CMS electromagnetic calorimeter. The results of the search are consistent with the background prediction and are interpreted using a gauge-mediated supersymmetry breaking reference model with a gluino next-to-lightest supersymmetric particle. In this model, gluino masses up to 2100, 2500, and 1900\GeV are excluded at 95\% confidence level for proper decay lengths of 0.3, 1, and 100\unit{m}, respectively. These are the best limits to date for such massive gluinos with proper decay lengths greater than $\sim$0.5\unit{m}.}

\hypersetup{
    pdftitle={Search for long-lived particles using nonprompt jets and missing transverse momentum with proton-proton collisions at sqrt(s)=13 TeV},
  pdfauthor={CMS Collaboration},
  pdfsubject={CMS, long-lived particles, displaced jets, supersymmetry, SUSY, GMSB},
  pdfkeywords={Exotica, Long-lived particles}
}

\maketitle
\section{Introduction}
\label{sec:intro}

A large number of models for physics beyond the standard model predict long-lived particles
that may be produced at the CERN LHC and decay into final states containing
jets with missing transverse momentum, \ptmiss~\cite{zhenSummary}. These models include
supersymmetry (SUSY) with gauge-mediated SUSY breaking (GMSB)~\cite{Giudice:1998bp},
split and stealth SUSY~\cite{ArkaniHamed:2004fb, Giudice:2004tc,Fan:2011yu}, and hidden valley models~\cite{Strassler:2006im}.
The \ptmiss may arise from a stable neutral weakly interacting particle in the final state
or from a heavy neutral long-lived particle that decays outside the detector.

The timing capabilities of the CMS electromagnetic calorimeter (ECAL)~\cite{ecal_tdr}
are used to identify nonprompt or ``delayed" jets produced by the displaced decays of heavy long-lived particles
within the ECAL volume or within the tracking volume bounded by the ECAL. The delay is expected to be a few ns for a TeV scale particle that travels
$\sim$1\unit{m} before decaying. A representative GMSB model is used
as a benchmark to quantify the sensitivity of the search.
In this model, pair-produced long-lived gluinos each decay
into a gluon, which forms a jet, and a gravitino, which escapes the detector causing significant
\ptmiss in the event. A diagram showing the benchmark model is shown in Fig.~\ref{fig:gluinoGMSB} (\cmsLeft).

\begin{figure}[!htb]
\centering
    \includegraphics[width=0.49\textwidth]{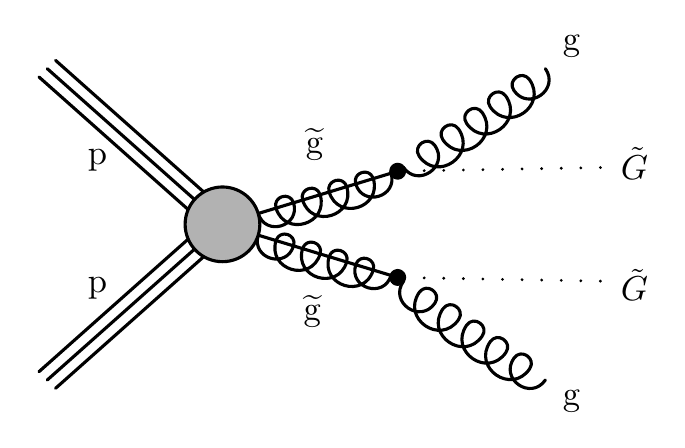}
    \includegraphics[width=0.49\textwidth]{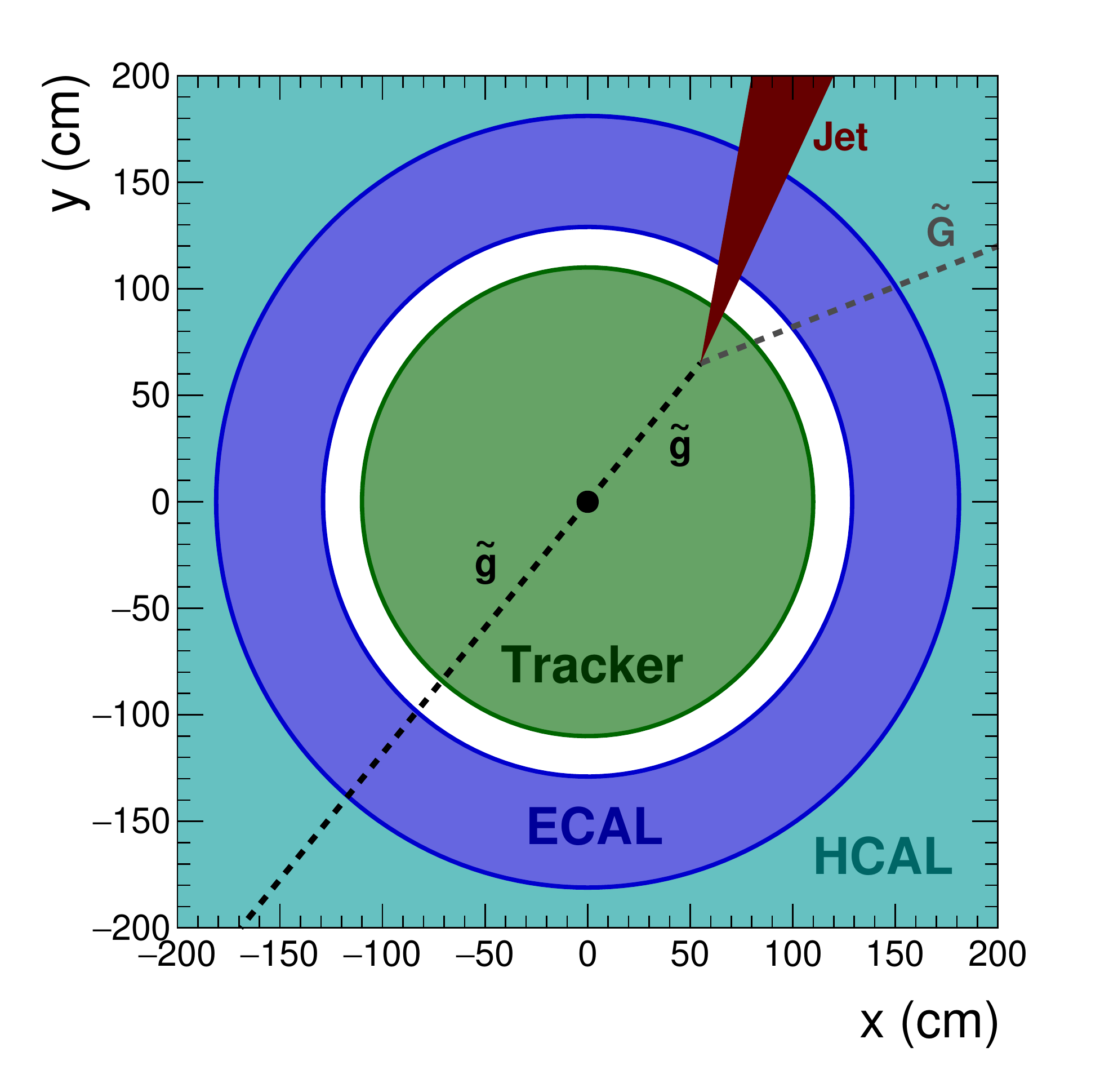}
    \caption{\label{fig:gluinoGMSB}
    Diagram showing the GMSB signal model (\cmsLeft), and diagram of a typical event (\cmsRight), expected to pass the signal region selection. The event has delayed energy depositions in the calorimeters but no tracks from a primary vertex.}
\end{figure}

There have been multiple searches for long-lived particles decaying to jets
by the ATLAS~\cite{ATLASTDR}, CMS~\cite{CMSTDR} and LHCb~\cite{LHCbTDR} Collaborations at $\sqrt{s}=7\TeV$, $\sqrt{s}=8\TeV$ and $\sqrt{s}=13\TeV$~\cite{Khachatryan:2010uf,
Aad:2012zn, Aad:2013gva, Khachatryan:2015jha, Aad:2015rba, Aaboud:2016dgf, Aaboud:2017iio,Sirunyan2018,Aaboud:2019opc,
displaced-dijets,Sirunyan:2018pwn,Aaij2017,ATLAS:2012av,Aad:2015uaa,Aaboud:2018aqj}. The use of calorimeter timing has so far been limited to searches targeting displaced photons at
$\sqrt{s}=8\TeV$~\cite{CMS-displaced-photons,ATLAS-displaced-photons}.
The present study represents the first application of ECAL timing to a search for
nonprompt jets from long-lived particle decays. This technique allows the reduction of background contributions
to the few event level, while retaining high efficiency for signal signatures of
one or more displaced jets and \ptmiss in the final state.
As detailed in Ref.~\cite{Liu:2018wte}, this approach brings
significant new sensitivity to long-lived particle searches. A diagram of a characteristic
event targeted by this analysis is shown in Fig.~\ref{fig:gluinoGMSB} (\cmsRight). Such an event would escape
reconstruction in a tracker-based search because of the difficulty in reconstructing tracks 
that originate from decay points separated from the primary vertex by more than $\sim$50\cm in the plane perpendicular to the beam axis.
There are two effects that contribute to the time delay of 
jets from the decay of heavy long-lived particles. First, 
the indirect path, composed of the initial long-lived particle 
and the subsequent jet trajectories, will be longer, and second, 
the long-lived particle will move with a lower velocity owing to 
its high mass. The latter is the dominant effect for the signal 
models considered in this analysis.

\section{The CMS detector}
\label{sec:detector}

The central feature of the CMS detector is a superconducting solenoid
of 6\unit{m} internal diameter, providing a magnetic field of
3.8\unit{T}. Within the solenoid volume are a silicon pixel and strip
tracker, a lead tungstate crystal ECAL,
and a brass and scintillator hadron calorimeter (HCAL), each composed
of a barrel and two endcap sections. Forward calorimeters extend the
pseudorapidity coverage provided by the barrel and endcap
detectors. The silicon tracker measures charged particles within the pseudorapidity range 
$\abs{\eta} < 2.5$. It consists of 1440 silicon pixel and 15\,148 silicon strip detector modules. 
For nonisolated particles with $1 < \pt < 10\GeV$, in the region $\abs{\eta} < 1.4$, the track resolutions 
are typically 1.5\% in \pt and 25--90 (45--150)\mum in the transverse (longitudinal) impact parameter~\cite{Chatrchyan:2014fea}.
The HCAL is segmented into individual calorimeter cells along pseudorapidity ($\eta$),
azimuth ($\phi$), and depth. The barrel muon system is composed of drift-tubes (DTs) and
resistive plate chambers (RPCs). These provide
high resolution hit positioning and timing to determine the muon trajectory.
The hits in the DTs are clustered into track segments, referred to as DT segments,
as detailed in Ref.~\cite{Chatrchyan:2009ih}. In the forward region, RPCs are used along with cathode strip chambers (CSCs), which
have greater resistance to the higher
radiation flux occurring along the beamline than DTs.
A more detailed description of the CMS detector, together with a definition of the
coordinate system used and the relevant kinematical variables, can be
found in Ref.~\cite{CMSTDR}.

The CMS ECAL consists of 75\,848 lead tungstate crystals, which provide coverage
in pseudorapidity $\abs{\eta}<1.48 $ in a barrel region (EB)
and $1.48<\abs{\eta}<3.00$ in two endcap regions (EE).
This analysis relies on the timing capabilities of the EB~\cite{ecal_tdr}.
The ECAL measures the energy of incoming electromagnetic particles through the scintillation light produced
in the lead tungstate crystals. Silicon avalanche photodiodes (APDs) are used as
photodetectors in the barrel region. These are capable of measuring the time
of incoming particles with a resolution as low as $\sim$200\unit{ps}
for energy deposits above 50\GeV~\cite{timing-note}.
Each ECAL crystal with an APD unit attached is referred to as an ECAL cell.

In the region $\abs{\eta}<1.74$, the HCAL cells have widths of 0.087 in $\eta$
and 0.087 in $\phi$. In the $\eta$--$\phi$ plane, and for $\abs{\eta}<1.48$, the HCAL cells
map on to $5{\times}5$ arrays of ECAL crystals to form calorimeter towers
projecting radially outwards from close to the nominal interaction point.
For $\abs{\eta}>1.74$, the coverage of the towers increases progressively
to a maximum of 0.174 in $\Delta \eta$ and $\Delta \phi$. Within each tower,
the energy deposits in ECAL and HCAL cells are summed to define the calorimeter
tower energies, subsequently used to provide the energies and directions of hadronic jets.

Events of interest are selected using a two-tiered trigger
system~\cite{Khachatryan:2016bia}. The first level, composed of custom
hardware processors, uses information from the calorimeters and muon
detectors to select events at a rate of around 100\unit{kHz} within a
time interval of less than 4\mus. The second level, known as the
high-level trigger (HLT), consists of a farm of processors running a version
of the full event reconstruction software optimized for fast
processing, and reduces the event rate to around 1\unit{kHz} before
data storage.

\section{Object and event reconstruction}
\label{sec:reconstruction}
The primary physics objects used in this analysis are jets reconstructed from the energy deposits
in the calorimeter towers, clustered using the anti-\kt algorithm~\cite{Cacciari:2008gp, Cacciari:2011ma}
with a distance parameter of 0.4. The contribution from each calorimeter tower is
assigned the coordinates of the tower and a momentum, the absolute value and the direction of which are found from the energy measured
in the tower assuming that the contributing particles originated at the center
of the detector. The raw jet energy is obtained from the sum of the
tower energies, and the raw jet momentum by the vectorial sum of the tower momenta, which are found from the energy measured in the tower.
The raw jet energies are then corrected to reflect a uniform relative
response of the calorimeter in $\eta$ and a calibrated absolute response in transverse momentum \pt~\cite{Khachatryan:2016kdb}.
Jets reconstructed using the CMS particle flow (PF) algorithm~\cite{CMS-PRF-14-001}
are not used in this analysis because nonprompt jets do not produce reliable information in the tracker and out-of-time energy
deposits are not included in the PF jet reconstruction.

All reconstructed vertices in the event,
consistent with originating from a proton-proton ($\Pp\Pp$) interaction, are considered to be primary vertices (PVs)~\cite{Chatrchyan:2014fea}.
Each track that is identified as originating from a PV
is associated with a jet if the separation of
the track from the jet axis $\Delta R=\sqrt{\smash[b]{(\Delta\eta)^2+(\Delta\phi)^2}}<0.4$,
where $\Delta\eta$ and $\Delta\phi$ represent the difference (in radians) between the jet axis and the track
in the pseudorapidity and in the azimuthal direction, respectively.

The jet timing is determined using all ECAL cells that satisfy $\Delta R<0.4$
between the jet axis and cell position, that exceed an energy threshold of 0.5\GeV
and that satisfy reconstruction quality criteria.
For each cell within the ECAL detector, the timing offset is defined such that a particle traveling at the speed of light from the center of the collision region to the cell position arrives at time zero.
Energy deposits with a recorded time that is either less than $-20\unit{ns}$ or greater than 20\unit{ns} are rejected, to remove events originating from preceding or following bunch collisions, respectively.
The time of the jet, \jt, is defined by the median cell time. The jet-based requirements
used to reject the dominant background sources, referred to as the signal jet requirements,
are detailed in Section~\ref{sec:selection}.

The missing transverse momentum vector, \ptvecmiss, used for this analysis is defined as
the projection on the plane perpendicular to the beams of the
negative vector sum of calorimeter
momenta deposits in an event satisfying reconstruction quality criteria 
chosen to reduce instrumental noise effects, 
but with no rejection of out-of-time ECAL cells.

\section{Data sets and simulated samples}
\label{sec:simulation}
The data sample was collected in 2016, 2017, and 2018 by
the CMS detector in $\Pp\Pp$ collisions at a center-of-mass energy of 13\TeV,
corresponding to an integrated luminosity of $137 \pm 3.3\fbinv$~\cite{CMS-PAS-LUM-17-001,CMS-PAS-LUM-17-004,CMS-PAS-LUM-18-002}.
The trigger required the events to satisfy $\ptmiss\textrm{ (trigger)}>120\GeV$. This is computed as the negative vector \ptvec sum of all
HLT PF candidates, which include out-of-time deposits~\cite{Sirunyan:2019kia}.

The search is interpreted using the GMSB signal model with samples produced
with gluino masses from 1000 to 3000\GeV, and proper decay lengths (\ctau) varying from
0.3 to 100\unit{m}. The gluino pair production cross sections
are determined at approximate-NNLO+NNLL order in $\alpha_\textrm{s}$
~\cite{Beenakker:1996ch, Kulesza:2008jb, Kulesza:2009kq,
  Beenakker:2009ha, Beenakker:2011fu, Borschensky:2014cia,Beenakker2016}. All other
SUSY particles, apart from the gravitino, are assumed to be heavy and
decoupled from the interaction. Signal samples are produced with
\PYTHIA 8.212~\cite{pythia}, and NNPDF3.1LO~\cite{nnpdf} is used for parton distribution function (PDF) modeling.
If a gluino is long-lived, it will have enough time to form a hadronic state,
an R-hadron~\cite{Fairbairn:2006gg, Kraan:2004tz,rhadorig},
which is simulated with \PYTHIA 8.212. For underlying event
modeling the CP2 tune is used~\cite{CP2}.

Systematic uncertainties in the modeling of the jet-based variables discussed in Section~\ref{sec:selection}
are derived using a simulated sample of jets produced through the strong interaction,
referred to as quantum chromodynamics (QCD) multijet events. This sample
is simulated with the \MGvATNLO 2.2.2~\cite{Alwall2014} event generator at
leading-order (LO) accuracy. This generator is interfaced with \PYTHIA 8.212 for
hadronization and fragmentation. The jets from the matrix element calculations are matched
to parton shower jets using the MLM algorithm~\cite{Alwall2014}.
The underlying event is modeled using the
CUETP8M1 (CP5) tune~\cite{CP2} for simulation
with NNPDF3.0NLO (NNPDF3.1NNLO)~\cite{nnpdf} used for PDF modeling for the 2016 (2017 and 2018)
detector operating conditions.

The description of the detector response is implemented using the
\GEANTfour~\cite{geant} package for all simulated processes.
To model the effect of additional $\Pp\Pp$ interactions within the same
bunch crossing (in-time pileup) or nearby bunch crossings (out-of-time pileup), minimum bias events generated with \PYTHIA are added to the simulated event sample, with a frequency distribution per bunch crossing weighted to match that observed in data.

\section{Event and object selection}
\label{sec:selection}
The selection criteria are optimized taking into account the principal background sources that produce delayed timing signals, which are detailed below.
\begin{itemize}
    \item ECAL time resolution tails: these tails affect the collisions of in-time (``core") bunches and arise from intercalibration uncertainties, crystal-dependent variations in scintillation rise time, loss of crystal transparency because of radiation, and run-by-run shifts associated with the readout electronics~\cite{timing-note}.
\item Electronic noise: electronic noise in the ECAL can cause individual cells to record deposits at arbitrary times, typically with low energies, and uncorrelated
    with surrounding cells.
\item Direct ionization in the APD: the traversal of a charged particle produces a signal that is $\sim$11\unit{ns} earlier than the signal from scintillation light. However, the ionization signal may arrive later if the associated charged particle travels back from the HCAL, or is associated with a later bunch crossing.
\item In-time pileup: additional $\Pp\Pp$ collisions in the same bunch crossing can produce particles with a spread in collision time and with varying flight paths, depending on the point of origin along the beam axis. These effects result in timing shifts of up to a few hundred ps.
\item Out-of-time pileup: additional $\Pp\Pp$ collisions in neighboring bunch crossings can result in deposits that are delayed by integer multiples of the bunch spacing (25\unit{ns}).
\item Satellite bunches: the LHC radiofrequency (RF) cavities operate at a frequency of 400 MHz, such that RF ``buckets" are separated by $\sim$2.5\unit{ns}. In order to achieve the desired bunch spacing, only one in ten of these buckets (separated by 25\unit{ns}) is filled. However, adjacent ``satellite" bunches may also contain protons at a level corresponding to $\mathcal{O}(10^{-5})$ times that of the main bunch.
\item Beam halo: collisions between beam protons and an LHC collimator~\cite{1590608} can result in muons that pass through the detector approximately parallel to the beam line. These ``beam halo" muons can deposit energy within the ECAL, causing an early signal if the beam halo is from the current or previous bunch or a delayed signal if the beam halo originates from a following bunch.
\item Cosmic ray muon hits: cosmic ray muons may cause deposits in the ECAL that occur at random times.
\end{itemize}

The events considered in this analysis as
including candidate long-lived particles are required to satisfy a
series of selections that define the signal region (SR). Each requirement
is chosen to be at least $\sim$90\% efficient for jets from the decay of a TeV scale long-lived
particle while allowing at least a factor $\sim$10 rejection
of the identified background process. In order to predict background contributions to the SR,
some of these requirements are inverted to enhance
particular background processes, as detailed in Section~\ref{sec:background}.

\subsection{Jet selection}
\label{sec:jetselection}

\subsubsection{Baseline jet selection}
All jets considered in this analysis must pass baseline \pt and $\eta$ requirements.
A requirement of $\pt>30\GeV$ is imposed to reduce contributions from pileup jets.
For the SR, further mitigation of pileup jets is achieved through selections detailed in
Section~\ref{sec:cleanselection}. The jets are required to satisfy $\abs{\eta}<1.48$ so that they are reconstructed in the EB.
The barrel requirement is made because the timing resolution is significantly
better in this region compared with the endcap~\cite{timing-note},
and jets of the targeted signal model are strongly peaked in the central $\eta$ region.

\subsubsection{Signal jet selection}
\label{sec:cleanselection}

The SR requirement on the jet time is $\jt>3\unit{ns}$.
The timing resolution improves for higher energy ECAL deposits before reaching a plateau~\cite{timing-note}.
A requirement on the ECAL energy component
of the jet of $\emj>20\GeV$ is applied as this threshold was found to optimize the timing resolution of the jets while ensuring high signal efficiency.

Jets from signal events are expected to have a large number of ECAL cells (\nec) hit,
while jets dominated by direct APD hits or ECAL noise often have a low number of cells hit.
A threshold of $\nec>25$ is applied to reject these background sources.

Jets from signal events will typically have similar energy depositions in
the ECAL and HCAL, while jets originating from noise or beam halo
typically have a small or zero HCAL energy component (\hmj). In order to reject such background sources,
jets are required to have a hadronic energy fraction $\hef=\hmj/(\emj+\hmj)>0.2$.
An additional requirement of $\hmj>50\GeV$ is
made to reject background contributions from noise and beam halo as well
as to ensure a well-measured hadronic component.

Signal jets typically have a small RMS in the time of the constituent cells (\jtrms) as all the component cells originate from the same delayed jet.
Jets that are significantly delayed because of contributions from uncorrelated noise often contain cells that are widely spread in time. In such cases the \jtrms will be correlated
with \jt, so a requirement is made on both $\jtrms<0.4\jt$ and $\jtrms<2.5\unit{ns}$.

Jets that originate from a PV and have a mismeasured time or originate from satellite bunch collisions
typically contain significant total momentum in tracks associated with their PV. The \pttf, defined as
the ratio of the total \pt of all
PV tracks matched to the jet ($\Delta R<0.5$) to the transverse calorimeter energy of the jet,
is used to select potential signal jets that do not originate from a PV. A requirement
of $\pttf<0.08$ is applied.

Beam halo muons will travel directly through the CSCs
before leaving energy deposits in the ECAL,
so the fraction of ECAL energy that can be associated with CSC hits
provides rejection of background contributions from beam halo. The ratio of the total energy of ECAL cells matched to
a CSC hit ($\Delta\phi<0.04$) to \emj, defined as \cscef, is used to discriminate
beam halo background contributions. A requirement of $\cscef<0.8$ is applied.

\subsection{Event selection}

The events are required to contain at least one jet satisfying the requirements outlined in
Section~\ref{sec:jetselection}. In addition, a requirement of $\ptmiss>300$\GeV is
applied to reject background contributions from multijet production from core and satellite bunch collisions.

The DT and RPC muon systems are used to reduce the background contribution from cosmic ray muons.
Signal events could also have deposits in the muon systems if the jets contain
muons, if there is ``punch-through" of jet constituents to the muon system,
or if the long-lived particle decays within the muon system. To mitigate the inefficiency
for signal events, only the DT segments and RPC hits with $r>560\cm$ (where $r$ is the transverse
radial distance to the interaction point) and RPC hits with $\abs{z}>600\cm$ (where $z$ is the distance along
the beamline to the interaction point) are considered. In order to reduce the effect of noise, DT segments and RPC hits
are required to be within $\Delta R<0.5$ of a DT segment with a hit. The maximal $\Delta \phi$
between such ``paired" DT segments and RPC hits is defined as \dtmaxdphi and \rpcmaxdphi, respectively.
Events satisfying $\dtmaxdphi>\pi/2$ or $\rpcmaxdphi>\pi/2$ are rejected to reduce the contribution of cosmic ray muon events.

Finally, events are required to satisfy a series of filters designed to
reject anomalous high-\ptmiss events, which can be due to a variety of reconstruction failures, 
detector malfunctions and backgrounds not arising from $\Pp\Pp$ collisions~\cite{Sirunyan:2019kia}. 
All SR requirements are summarized in Table~\ref{tab:signal_selections}.

\begin{table}[h!]
    \centering
    \topcaption{
	Summary of the requirements used to define the signal region.
    }
	\bgroup
	\def\arraystretch{1.2}
	\begin{tabular}{l}
	    \hline
	    \multicolumn{1}{c}{\textit{Baseline jet selection}}\\
	     $\abs{\eta}<1.48$ \\
	     $\pt>30\GeV$\\
	    \multicolumn{1}{c}{\textit{Signal jet selection}}\\
	     $\emj>20\GeV$\\
	     $\nec>25$ \\
	     $\hef>0.2$ and $\hmj>50\GeV$ \\
	     $\jtrmsfrac<0.4$ and $\jtrms<2.5\unit{ns}$\\
	     $\pttf<0.08$\\
	     $\cscef<0.8$\\
	     $\jt>3\unit{ns}$\\
	     \multicolumn{1}{c}{\textit{Event level selection}}\\
	     At least one signal jet\\
	     $\ptmiss>300\GeV$ \\
	     Quality filters\\
	     $\dtmaxdphi<\pi/2$ \\
	     $\rpcmaxdphi<\pi/2$ \\
	    \hline
	    \label{tab:signal_selections}
	\end{tabular}
	\egroup
\end{table}

\section{Background estimation}
\label{sec:background}
This section details the characterization of the dominant background
sources and the methods used to estimate residual contributions to the SR.
The background contributions are investigated by inverting the requirements on
the discriminating variables summarized in Table~\ref{tab:signal_selections} to define control regions (CRs)
enriched in particular background processes. There are three main background sources:
beam halo muons deposits, which typically have low \hef and large \cscef;
out-of-time jets from core and satellite bunch collisions, which have large \pttf;
and jets originating from cosmic ray muons, which have high \rpcdtmaxdphi and \jtrms.
The background sources are estimated from the CRs using methods that
rely on data. These predictions are tested using validation regions (VRs)
that do not overlap with the SRs to ensure they are unbiased. The agreement of the observation with
prediction in the VRs is used to estimate
systematic uncertainties in the prediction
in the SR. For jets in the CRs and VRs with $\abs{\jt}<3\unit{ns}$, the $\jtrmsfrac<0.4$ requirement is replaced
with a requirement of $\jtrms<1.2\unit{ns}$.

\subsection{Beam halo}
\label{sec:beam_halo_background}

\sloppy The beam halo contribution is estimated by measuring the pass/fail
ratio of the $\cscef>0.8$ requirement for events with $\hef<0.2$
and applying it to the observed number of events with $\hef>0.2$. The prediction
is made without any requirement on \hmj and can therefore be considered
an upper limit on the contribution from the beam halo background contribution.

The VR for this prediction is defined by selecting events with $\jt<-2\unit{ns}$ and applying
all signal requirements except those on \cscef, \hef, and \hmj. To enhance the
contribution of beam halo events relative to the contributions from satellite
bunches and cosmic ray muons in the VR, the $\phi$ values of the jets are
required to be within 0.2 radians of 0 or $\pm\pi$. The correlation between \cscef and \hef in the VR is consistent with zero, meaning they
can be used to make an unbiased prediction. The prediction
from this method for the number of events passing signal thresholds on \cscef and \hef
in the VR is $0.02^{+0.06}_{-0.02}$ events, in agreement with the 0
events observed.

The level of agreement between prediction and observation in the VR
is used to derive a systematic uncertainty
in the prediction. The slope of a linear fit to the pass/fail
ratio of the $\cscef>0.8$ requirement as a function of \hef is found to
be consistent with zero. The uncertainty is then propagated to the region
with $\cscef>0.8$ and $\hef>0.2$.
The final prediction for the SR is
$0.02^{+0.06}_{-0.02}\stat^{+0.05}_{-0.01}\syst$ events.

\subsection{Core and satellite bunch background prediction}

The core and satellite bunch background contribution is estimated by measuring the
pass/fail ratio of the requirement $\pttf<0.08$ for events with $1<\jt<3\unit{ns}$
and applying it to the observed number of events with $\jt>3\unit{ns}$ and $\pttf>0.08$.
Two VRs are defined to verify the prediction of the
satellite bunch and timing tail background contributions.

The first VR is selected to contain events with
$\jt<-1\unit{ns}$ and passing all signal requirements except for that on \pttf.
The pass/fail ratio of the $\pttf<0.08$ requirement is measured for events
with $-3<\jt<-1\unit{ns}$ and applied to the number of
events with $\jt<-3\unit{ns}$ and $\pttf>0.08$. The upper bound on \jt ensures the sample is enriched with jets in the tail of the
\jt distribution. The correlation between the variables in the VR
is confirmed to be consistent with zero, which allows an unbiased prediction to be made.
The prediction from this method for the number
of events passing $\jt<-3\unit{ns}$ and $\pttf <0.08$ is $0.09^{+0.2}_{-0.06}$ events,
to be compared with 1 observed event. The event passing selection
has no paired RPC or DT hits and is therefore unlikely to originate from a cosmic ray muon.
The compatibility with expectation is within two standard deviations,
however, to ensure the prediction is unbiased, a further validation is carried out.
The requirement of $\ptmiss>300\GeV$ is inverted and the prediction repeated. The
events must still satisfy the $\ptmiss\textrm{ (trigger)}>120\GeV$ requirement.
The number of events satisfying $\jt<-3\unit{ns}$ and $\pttf <0.08$ is predicted
to be $1.95\pm 0.29$ events, to be compared with 1 event observed. 
As the validation with $\ptmiss<300\GeV$ 
probes a similar phase space to the validation with $\ptmiss>300\GeV$, 
but with a significantly increased number of events, an excess 
due to a systematic effect would be enhanced. The observation in the
region with $\jt<-3\unit{ns}$ and $\pttf <0.08$, for $\ptmiss>300\GeV$,
is therefore considered to be consistent with a statistical fluctuation.

A second VR is defined using events with $1<\jt<3\unit{ns}$.
The pass/fail ratio of the $\pttf<0.08$ requirement is measured for events
with $1<\jt<2\unit{ns}$ and applied to the number of
events with $2<\jt<3\unit{ns}$ and $\pttf>0.08$.
The estimation from this method for the number
of events passing $2<\jt<3\unit{ns}$ and $\pttf <0.08$ is $0.03^{+0.08}_{-0.03}$ events, in agreement with the 0
events observed.

The prediction for the SR relies on using the efficiency of the \pttf requirement of events
with $1<\jt<3\unit{ns}$ to predict the efficiency of the \pttf requirement for $\jt>3\unit{ns}$.
Because of differences in the reconstruction of the calorimeter energy and tracker \pt, this
efficiency may be expected to have some small time dependence. In order to measure
any such \jt dependence and derive an associated systematic uncertainty, a data sample with
the offline \ptmiss requirement inverted (but passing trigger requirements)
and $\jt>2\unit{ns}$ is used. The region of $\pttf<0.08$
is not included to avoid contamination from cosmic ray or beam halo muon deposits. The slope of
a linear fit to the pass/fail ratio of a looser requirement of $\pttf<0.5$ against \jt
is consistent with zero.
As for the beam halo prediction, the uncertainty from the fit is
propagated to the region with $\jt>3\unit{ns}$ and $\pttf>0.08$. The final
prediction for the core and satellite bunch background contribution is
$0.11^{+0.09}_{-0.05}\stat^{+0.02}_{-0.02}\syst$ events.

\subsection{Cosmic ray events}

The discriminating variables used for the cosmic background prediction are the $\jtrms$
of the jet and the larger of \dtmaxdphi and \rpcmaxdphi, labeled as \rpcdtmaxdphi.
The pass/fail ratio of the $\jtrms<2.5\unit{ns}$ requirement is measured
for events with $\rpcdtmaxdphi>\pi/2$ and applied to events with
$\rpcdtmaxdphi<\pi/2$.
Cosmic ray muons that radiate a photon via bremsstrahlung while passing 
through the HCAL will typically deposit significant energy in a single
isolated cell. The HCAL noise rejection quality filters are designed to reject events containing such
isolated deposits, thus inverting these filters, with all
other requirements applied, provides a validation
region enriched in events with cosmic ray muons.

The correlation between \jtrms and \rpcdtmaxdphi in the validation sample is consistent with zero, allowing them to be used to make an unbiased prediction. The estimation in
the VR for the number of events passing signal thresholds in \jtrms and \rpcdtmaxdphi
is $1.1^{+1.9}_{-1.1}$ events, in agreement with the 1 event observed.
A systematic uncertainty in the SR prediction is derived from the statistical uncertainty in the VR.
The final prediction in the SR is $1.0^{+1.8}_{-1.0}\stat^{+1.8}_{-1.0}\syst$ events.

\subsection{Background summary}
The estimated background yields and uncertainties are summarized in Table~\ref{tab:bkgsum}.
The total background prediction is $1.1^{+2.5}_{-1.1}$ events.

\begin{table}[h!]
    \centering
    \topcaption{
	Summary of the estimated number of background events.	
    }
	\bgroup
	\def\arraystretch{1.2}
	\begin{tabular}{lc}
	    \hline
	    Background source & Events predicted \\
	    \hline
	    Beam halo muons &$0.02^{+0.06}_{-0.02}\stat^{+0.05}_{-0.01}\syst$\\
	    Core and satellite& \multirow{2}{*}{$0.11^{+0.09}_{-0.05}\stat^{+0.02}_{-0.02}\syst$}\\
        bunch collisions &\\
	    Cosmic ray muons & $1.0^{+1.8}_{-1.0}\stat^{+1.8}_{-1.0}\syst$ \\
	    \hline
	    \label{tab:bkgsum}
	\end{tabular}
	\egroup
\end{table}

\section{Results and interpretation}
\label{sec:interpretation}

\begin{figure}[!htb]
\centering
    \includegraphics[width=\cmsFigWidth]{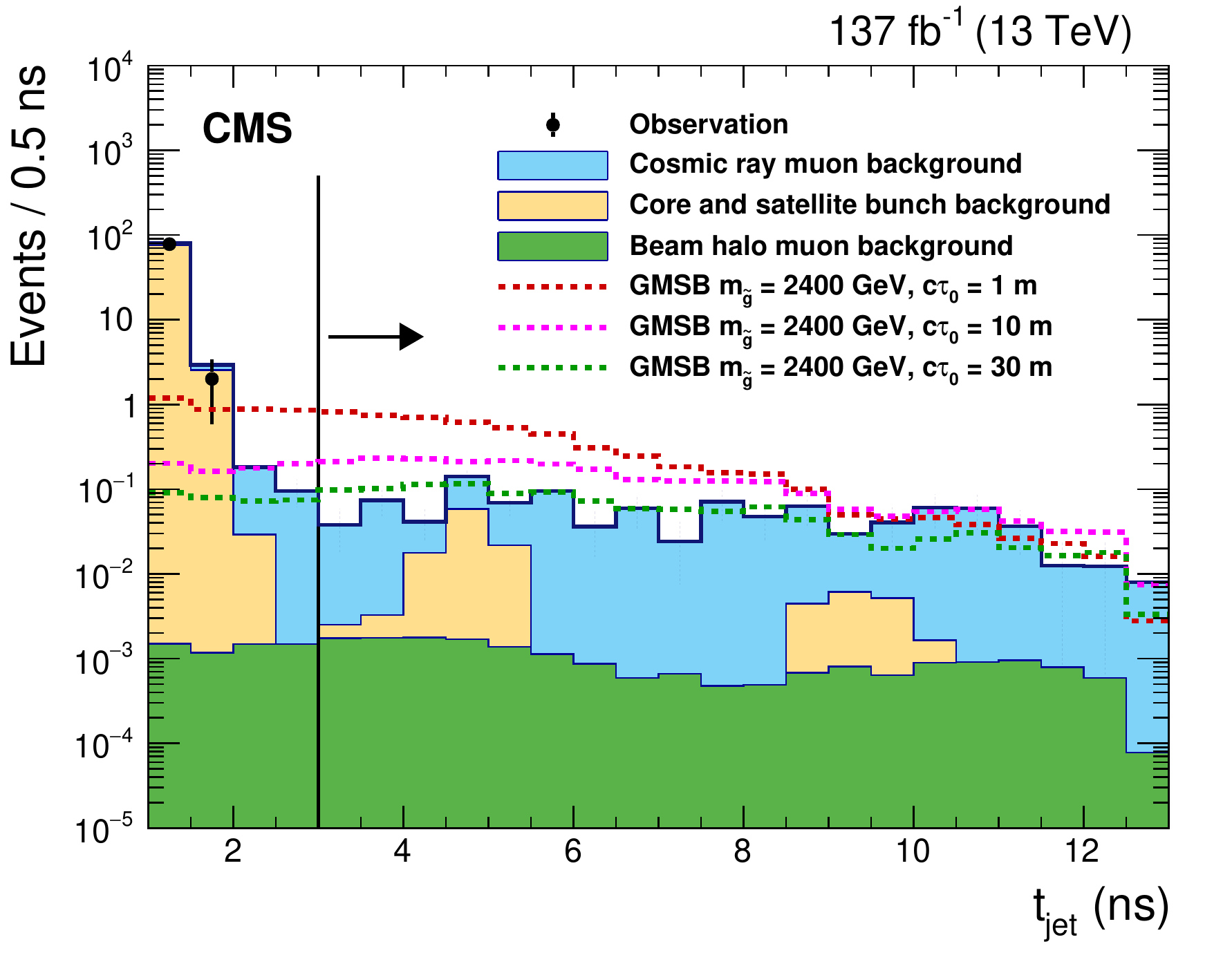}
  \caption{\label{fig:backgroundTemplate} The timing distribution of the background sources predicted to contribute to
  the signal region, compared to those for a representative signal model. The time is defined by the jet
    in the event with the largest \jt passing the relevant selection. The distributions for the major background sources
    are taken from control regions and normalized to the predictions detailed in Section~\ref{sec:background}.
    The observed data is shown by the black points. No events are observed in data for $\jt>3\unit{ns}$ (indicated with a vertical black line).}
\end{figure}

Figure~\ref{fig:backgroundTemplate} shows the timing distribution for events with jets passing all the SR requirements.
The distributions for the major background sources are taken from control regions and normalized to the
predictions detailed in Section~\ref{sec:background}. These distributions are shown for illustration 
only and are not used for the statistical interpretation.
The overall background prediction for the SR is $1.1^{+2.5}_{-1.1}$ events, which is consistent with the observation of 0 events.

The model used for the interpretation is the GMSB SUSY model in which gluinos are pair
produced and form R-hadrons. The long-lived gluinos then decay to a gluon and gravitino
producing a delayed jet and \ptmiss.

The trigger efficiency for the simulated samples is evaluated from an
emulation. The inefficiency due to the \ptmiss trigger requirement
ranges from $\sim$5 to $\sim$15\% for $\ctau=1$ and 10\unit{m}, respectively.
The trigger emulation is validated with data using an independent sample collected
with a single muon reference trigger.

\begin{figure}[!hbt]
    \centering
      \includegraphics[width=\cmsFigWidth]{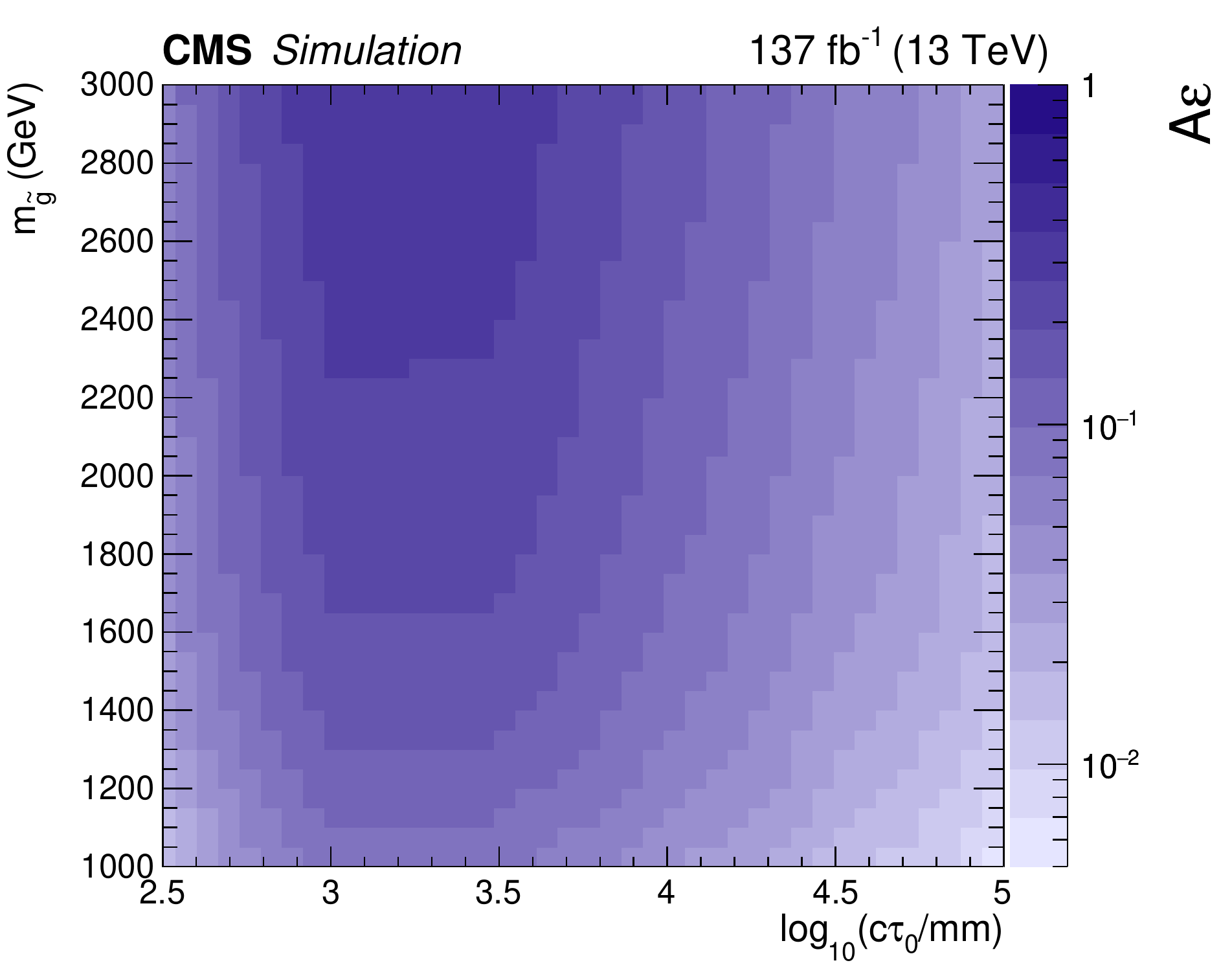}
   \caption{
   The product, \ate, of the acceptance and efficiency in the \ctau vs. \mg plane for the GMSB model, after all requirements.
   \label{fig:effGluinoGMSB} }
\end{figure}

The product of the experimental acceptance and efficiency (\ate), shown in Fig.~\ref{fig:effGluinoGMSB},
is evaluated independently for each model point, defined in terms of the gluino mass (\mg) and proper decay length.
The efficiency is maximized for high gluino masses and for a range in \ctau bounded
by the requirements that the gluino must have sufficient lifetime for its decay products
to pass the $\jt>3\unit{ns}$ requirement and that the gluino must decay before or within the ECAL.
For a gluino model with $\mg=2400\GeV$ the efficiency is highest (up to $\sim$35\%) for the range $1<\ctau<10\unit{m}$.
The efficiency is larger for higher masses because of the increased \ptmiss in the event and the reduced velocity of the gluino.

Interactions of the R-hadrons
with the detector lead to signatures exploited by searches for heavy stable charged particles
and, in order to maintain model independence, are not considered for the interpretation of this analysis.
However, the impact of such interactions was evaluated for two benchmark signal points,
$\mg=1500\GeV$ and $\ctau=1\unit{m}$, and $\mg=1500\GeV$ and $\ctau=10\unit{m}$, using 
the ``cloud" model of R-hadron/matter interactions~\cite{Kraan:2004tz,Mackeprang:2006gx}, which assumes that the R-hadron is
surrounded by a cloud of colored, light constituents that interact during scattering.
The fraction of \PSg which hadronize to a neutral \PSg-gluon state was taken to be 0.1.
Compared to non-interacting R-hadrons, the relative
reduction in selection efficiency for both benchmark signal points was found to be $\sim$15\% with the largest
effect being on the \pttf and \rpcdtmaxdphi requirements.

\subsection{Signal systematic uncertainties}

\begin{table}
    \centering
    \topcaption{
    The derived uncertainty in the product, \ate, of the acceptance and efficiency from the modeling of the variables discussed in
    Section~\ref{sec:cleanselection}, for a representative model with $\mg=2400\GeV$.
    }
    \bgroup
	\def\arraystretch{1.2}
	\begin{tabular}{lcc}
	    \hline
	    Variable & \multicolumn{2}{c}{Derived uncertainty (\%)}\\
		     & $\ctau=1\unit{m}$&$\ctau=10\unit{m}$ \\
	    \hline
	    \pttf  &0.01  & 0.03    \\
	    \nec   &3.2   &  4.2    \\
	    \hef   &2.8  &   2.5    \\
	    \cscef &0.9  &   0.9    \\
	    \jtrms &22 &  15    \\
	    \hline
	\end{tabular}
    \egroup
	    \label{tab:eff_comp}
\end{table}

In order to evaluate systematic uncertainties in the modeling of the variables used to select signal jets
(defined in Section~\ref{sec:cleanselection}), the corresponding distributions for events from the multijet simulation
are compared with data. For each variable, the threshold used for the selection is varied
in the simulation to match the efficiency measured in data. The change in acceptance from
this variation is shown for each of the jet-based variables in Table~\ref{tab:eff_comp},
using an example model point.
This variation is taken as a systematic uncertainty in the signal model acceptance.
In addition, the variation in \jtrms is propagated to \jtrmsfrac.

In addition to the uncertainty in the modeling of the variables used to select signal jets,
the systematic uncertainties in the signal \ate are summarized below.

\begin{itemize}
\item Integrated luminosity: 2.5\%~\cite{CMS-PAS-LUM-17-001}, 2.3\%~\cite{CMS-PAS-LUM-17-004}, and 2.5\%~\cite{CMS-PAS-LUM-18-002} uncorrelated uncertainties for the 2016, 2017, and 2018 data taking periods, respectively.
\item Trigger inefficiency: typically 5--15\%.
\item Limited simulated sample size: up to $\sim$10\%, depending on SR \ate.
\item Pileup reweighting: 4.6\% uncertainty in the total inelastic $\Pp\Pp$ cross section~\cite{inelast}, which corresponds to an uncertainty
    in the SR \ate of 1--5\%.
\item Jet energy resolution/scale: a 1--5\% percent uncertainty~\cite{Khachatryan:2016kdb}.
\end{itemize}

\subsection{Interpretation}

Under the signal plus background hypothesis, a modified frequentist approach is used
to determine observed upper limits at 95\% confidence level (\CL)
on the cross section ($\sigma$) to produce a pair
of gluinos, each decaying with 100\% branching fraction
to a gluon and a gravitino, as a function of \mg and \ctau.
The approach uses the LHC-style profile likelihood ratio
as the test statistic~\cite{CMS-NOTE-2011-005} and the \CLs
criterion~\cite{junk, CLsTechnique}. The expected and observed upper limits are evaluated
through the use of pseudodata sets. Potential signal contributions to event counts in
the SR and CRs are taken into consideration.

Figure~\ref{fig:absLimit} shows the observed upper limit on $\sigma$ as a function of lifetime and gluino mass for the GMSB model.
Gluino masses below 2100\GeV are excluded at 95\% confidence level for \ctau between 0.3 and 30\unit{m}. The dependence of the expected and observed upper limit as a function of \ctau is shown in Fig.~\ref{fig:limGMSB}
for $\mg=2400\GeV$. The observed limit is compared to the results of the CMS displaced jet search~\cite{displaced-dijets}, based on a data sample with integrated luminosity of $36.1\fbinv$, showing the complementary coverage. These results extend the reach beyond previous searches for models with jets and significant \ptmiss in the final state for $\ctau \gtrsim0.5\unit{m}$~\cite{displaced-dijets,Sirunyan:2018pwn,Aaboud:2017iio}.

\section{Summary}
\label{sec:summary}
An inclusive search for long-lived particles has been presented, based on a data sample of proton-proton collisions collected
at $\sqrt{s}=13\TeV$ by the CMS experiment, corresponding to an integrated luminosity of 137\fbinv. The search uses the timing of energy deposits
in the electromagnetic calorimeter to select delayed jets from the decays of heavy long-lived particles,
with residual background contributions estimated using measurements in control regions in the data.
The results are interpreted using the gluino gauge-mediated supersymmetry breaking signal model and
gluino masses up to 2100, 2500, and $1900\GeV$ are excluded at
$95\%$ confidence level for proper decay lengths of 0.3, 1, and $100\unit{m}$, respectively.
The reach for models that predict significant missing transverse momentum in the final state
 is significantly extended beyond all previous searches, for proper decay lengths greater than $\sim$0.5\unit{m}.

\begin{figure}[hbt!]
\centering
      \includegraphics[width=\cmsFigWidth]{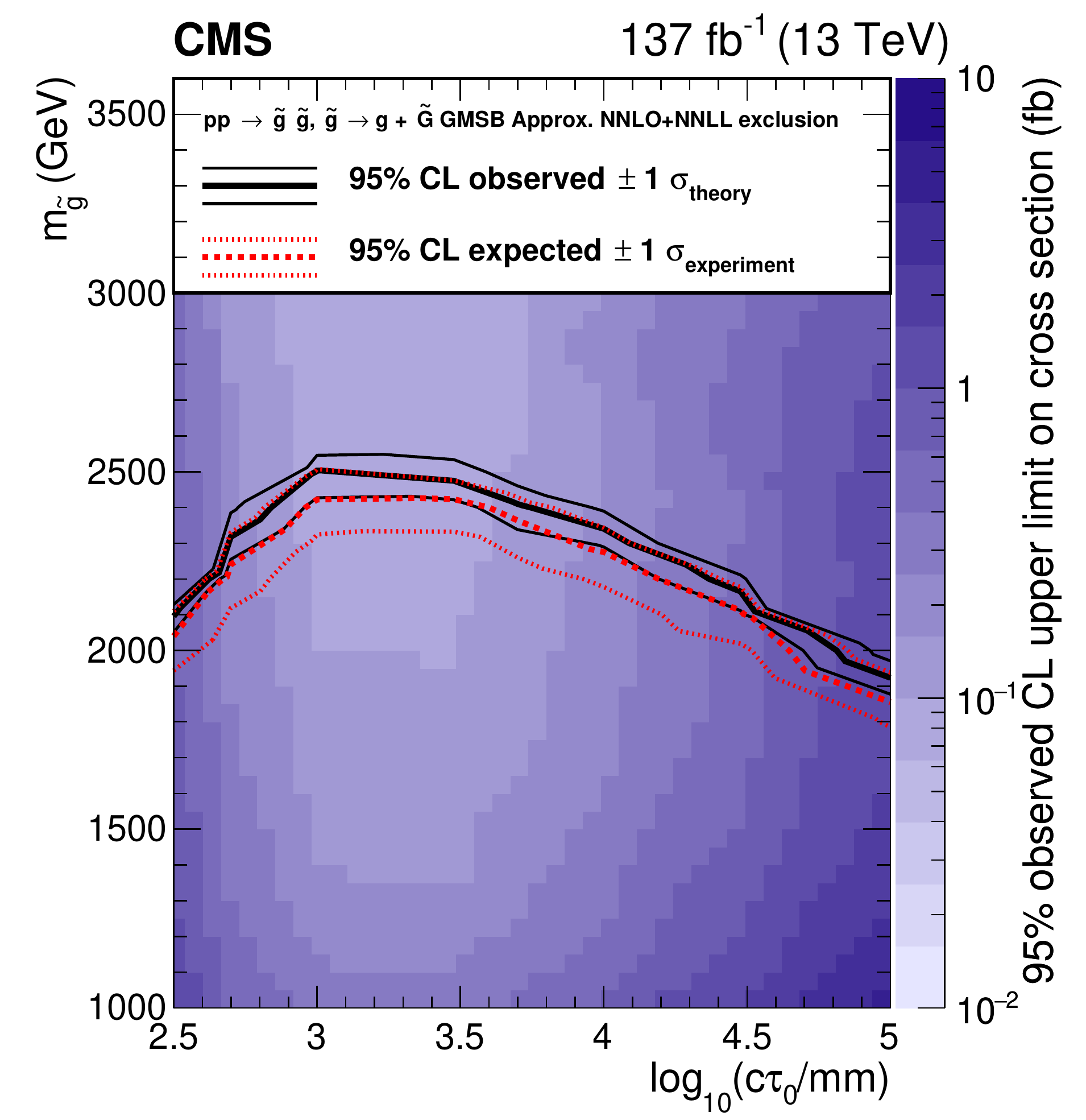}
       \caption{The observed upper limits at 95\% \CL for the gluino pair production cross section in the GMSB model, shown in the plane of \mg and \ctau. A branching fraction of 100\% for the gluino decay to a gluon and a gravitino is assumed. The area below the thick black curve represents the observed exclusion region,  while the dashed red lines indicate the expected limits and their $\pm 1$ standard deviation ranges.  The thin black lines show the effect of the theoretical uncertainties on the signal cross section\label{fig:absLimit}.}
\end{figure}

\begin{figure}[hbt!]
\centering
      \includegraphics[width=\cmsFigWidth]{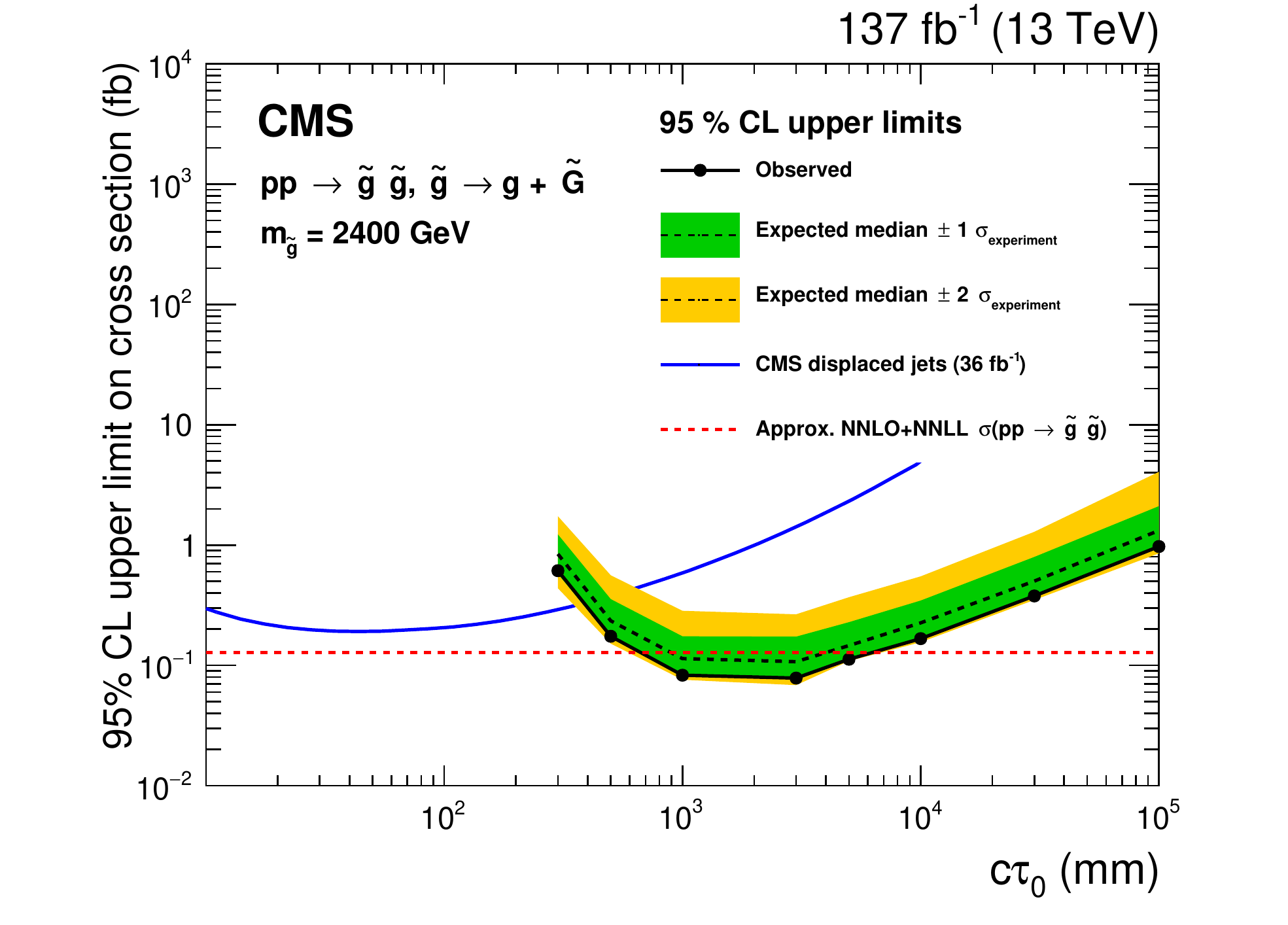}
   \caption{
       The observed and expected upper limits at 95\% CL on the gluino pair production cross section for
       a gluino GMSB model with $\mg=2400\GeV$. The one (two) standard deviation variation in the expected limit is shown in the inner green (outer yellow) band. The blue solid line shows
       the observed limit obtained by the CMS displaced jet search~\cite{displaced-dijets}\label{fig:limGMSB}. }
\end{figure}

\begin{acknowledgments}
We congratulate our colleagues in the CERN accelerator departments for the excellent performance of the LHC and thank the technical and administrative staffs at CERN and at other CMS institutes for their contributions to the success of the CMS effort. In addition, we gratefully acknowledge the computing centers and personnel of the Worldwide LHC Computing Grid for delivering so effectively the computing infrastructure essential to our analyses. Finally, we acknowledge the enduring support for the construction and operation of the LHC and the CMS detector provided by the following funding agencies: BMBWF and FWF (Austria); FNRS and FWO (Belgium); CNPq, CAPES, FAPERJ, FAPERGS, and FAPESP (Brazil); MES (Bulgaria); CERN; CAS, MoST, and NSFC (China); COLCIENCIAS (Colombia); MSES and CSF (Croatia); RPF (Cyprus); SENESCYT (Ecuador); MoER, ERC IUT, PUT and ERDF (Estonia); Academy of Finland, MEC, and HIP (Finland); CEA and CNRS/IN2P3 (France); BMBF, DFG, and HGF (Germany); GSRT (Greece); NKFIA (Hungary); DAE and DST (India); IPM (Iran); SFI (Ireland); INFN (Italy); MSIP and NRF (Republic of Korea); MES (Latvia); LAS (Lithuania); MOE and UM (Malaysia); BUAP, CINVESTAV, CONACYT, LNS, SEP, and UASLP-FAI (Mexico); MOS (Montenegro); MBIE (New Zealand); PAEC (Pakistan); MSHE and NSC (Poland); FCT (Portugal); JINR (Dubna); MON, RosAtom, RAS, RFBR, and NRC KI (Russia); MESTD (Serbia); SEIDI, CPAN, PCTI, and FEDER (Spain); MOSTR (Sri Lanka); Swiss Funding Agencies (Switzerland); MST (Taipei); ThEPCenter, IPST, STAR, and NSTDA (Thailand); TUBITAK and TAEK (Turkey); NASU and SFFR (Ukraine); STFC (United Kingdom); DOE and NSF (USA).

\hyphenation{Rachada-pisek} Individuals have received support from the Marie-Curie program and the European Research Council and Horizon 2020 Grant, contract Nos.\ 675440, 752730, and 765710 (European Union); the Leventis Foundation; the A.P.\ Sloan Foundation; the Alexander von Humboldt Foundation; the Belgian Federal Science Policy Office; the Fonds pour la Formation \`a la Recherche dans l'Industrie et dans l'Agriculture (FRIA-Belgium); the Agentschap voor Innovatie door Wetenschap en Technologie (IWT-Belgium); the F.R.S.-FNRS and FWO (Belgium) under the ``Excellence of Science -- EOS" -- be.h project n.\ 30820817; the Beijing Municipal Science \& Technology Commission, No. Z181100004218003; the Ministry of Education, Youth and Sports (MEYS) of the Czech Republic; the Lend\"ulet (``Momentum") Program and the J\'anos Bolyai Research Scholarship of the Hungarian Academy of Sciences, the New National Excellence Program \'UNKP, the NKFIA research grants 123842, 123959, 124845, 124850, 125105, 128713, 128786, and 129058 (Hungary); the Council of Science and Industrial Research, India; the HOMING PLUS program of the Foundation for Polish Science, cofinanced from European Union, Regional Development Fund, the Mobility Plus program of the Ministry of Science and Higher Education, the National Science Center (Poland), contracts Harmonia 2014/14/M/ST2/00428, Opus 2014/13/B/ST2/02543, 2014/15/B/ST2/03998, and 2015/19/B/ST2/02861, Sonata-bis 2012/07/E/ST2/01406; the National Priorities Research Program by Qatar National Research Fund; the Ministry of Science and Education, grant no. 3.2989.2017 (Russia); the Programa Estatal de Fomento de la Investigaci{\'o}n Cient{\'i}fica y T{\'e}cnica de Excelencia Mar\'{\i}a de Maeztu, grant MDM-2015-0509 and the Programa Severo Ochoa del Principado de Asturias; the Thalis and Aristeia programs cofinanced by EU-ESF and the Greek NSRF; the Rachadapisek Sompot Fund for Postdoctoral Fellowship, Chulalongkorn University and the Chulalongkorn Academic into Its 2nd Century Project Advancement Project (Thailand); the Welch Foundation, contract C-1845; and the Weston Havens Foundation (USA).
\end{acknowledgments}
\bibliography{auto_generated}

\providecommand{\href}[2]{#2}\begingroup\raggedright\begin{thebibliography}{10}%
\makeatletter
\providecommand{\hrefCMSnoop }[0]{\@secondoftwo}%
\makeatother
\providecommand{\doi}{\texttt{doi:}\begingroup \urlstyle{tt}\Url}

\bibitem{zhenSummary}
\hrefCMSnoop {}{Z.~Liu and B.~Tweedie, ``{The fate of long-lived superparticles
  with hadronic decays after LHC run 1}'',} \textit{ JHEP} \textbf{ 06} (2015)
  042,
  \href{http://dx.doi.org/10.1007/JHEP06(2015)042}{\doi{10.1007/JHEP06(2015)042}},
\href{http://www.arXiv.org/abs/1503.05923}{\texttt{arXiv:1503.05923}}.

\bibitem{Giudice:1998bp}
\hrefCMSnoop {}{G.~F. Giudice and R.~Rattazzi, ``{Theories with gauge mediated
  supersymmetry breaking}'',} \textit{ Phys. Rept.} \textbf{ 322} (1999) 419,
  \href{http://dx.doi.org/10.1016/S0370-1573(99)00042-3}{\doi{10.1016/S0370-1573(99)00042-3}},
\href{http://www.arXiv.org/abs/hep-ph/9801271}{\texttt{arXiv:hep-ph/9801271}}.

\bibitem{ArkaniHamed:2004fb}
\hrefCMSnoop {}{N.~Arkani-Hamed and S.~Dimopoulos, ``{Supersymmetric
  unification without low energy supersymmetry and signatures for fine-tuning
  at the LHC}'',} \textit{ JHEP} \textbf{ 06} (2005) 073,
  \href{http://dx.doi.org/10.1088/1126-6708/2005/06/073}{\doi{10.1088/1126-6708/2005/06/073}},
\href{http://www.arXiv.org/abs/hep-th/0405159}{\texttt{arXiv:hep-th/0405159}}.

\bibitem{Giudice:2004tc}
\hrefCMSnoop {}{G.~F. Giudice and A.~Romanino, ``{Split supersymmetry}'',}
  \textit{ Nucl. Phys. B} \textbf{ 699} (2004) 65,
  \href{http://dx.doi.org/10.1016/j.nuclphysb.2004.08.001}{\doi{10.1016/j.nuclphysb.2004.08.001}},
  \href{http://www.arXiv.org/abs/hep-ph/0406088}{\texttt{arXiv:hep-ph/0406088}}.
[Erratum: \DOI{10.1016/j.nuclphysb.2004.11.048}].

\bibitem{Fan:2011yu}
\hrefCMSnoop {}{J.~Fan, M.~Reece, and J.~T. Ruderman, ``{Stealth
  supersymmetry}'',} \textit{ JHEP} \textbf{ 11} (2011) 012,
  \href{http://dx.doi.org/10.1007/JHEP11(2011)012}{\doi{10.1007/JHEP11(2011)012}},
\href{http://www.arXiv.org/abs/1105.5135}{\texttt{arXiv:1105.5135}}.

\bibitem{Strassler:2006im}
\hrefCMSnoop {}{M.~J. Strassler and K.~M. Zurek, ``{Echoes of a hidden valley
  at hadron colliders}'',} \textit{ Phys. Lett. B} \textbf{ 651} (2007) 374,
  \href{http://dx.doi.org/10.1016/j.physletb.2007.06.055}{\doi{10.1016/j.physletb.2007.06.055}},
\href{http://www.arXiv.org/abs/hep-ph/0604261}{\texttt{arXiv:hep-ph/0604261}}.

\bibitem{ecal_tdr}
\href {https://cds.cern.ch/record/349375}{{CMS Collaboration}, ``{The CMS
  electromagnetic calorimeter project: Technical Design Report}'',} Technical
  Report CERN-LHCC-97-033, 1997.

\bibitem{ATLASTDR}
\hrefCMSnoop {}{{ATLAS Collaboration}, ``{The ATLAS Experiment at the CERN
  Large Hadron Collider}'',} \textit{ JINST} \textbf{ 3} (2008) S08003,
\href{http://dx.doi.org/10.1088/1748-0221/3/08/S08003}{\doi{10.1088/1748-0221/3/08/S08003}}.

\bibitem{CMSTDR}
\hrefCMSnoop {}{{CMS Collaboration}, ``{The CMS Experiment at the CERN LHC}'',}
  \textit{ JINST} \textbf{ 3} (2008) S08004,
\href{http://dx.doi.org/10.1088/1748-0221/3/08/S08004}{\doi{10.1088/1748-0221/3/08/S08004}}.

\bibitem{LHCbTDR}
\hrefCMSnoop {}{{LHCb Collaboration}, ``{The LHCb Detector at the LHC}'',}
  \textit{ JINST} \textbf{ 3} (2008) S08005,
\href{http://dx.doi.org/10.1088/1748-0221/3/08/S08005}{\doi{10.1088/1748-0221/3/08/S08005}}.

\bibitem{Khachatryan:2010uf}
\hrefCMSnoop {}{{CMS Collaboration}, ``{Search for stopped gluinos in pp
  collisions at $\sqrt{s}=7\TeV$}'',} \textit{ Phys. Rev. Lett.} \textbf{ 106}
  (2011) 011801,
  \href{http://dx.doi.org/10.1103/PhysRevLett.106.011801}{\doi{10.1103/PhysRevLett.106.011801}},
\href{http://www.arXiv.org/abs/1011.5861}{\texttt{arXiv:1011.5861}}.

\bibitem{Aad:2012zn}
\hrefCMSnoop {}{{ATLAS Collaboration}, ``{Search for decays of stopped,
  long-lived particles from $7\TeV$ pp collisions with the ATLAS detector}'',}
  \textit{ Eur. Phys. J. C} \textbf{ 72} (2012) 1965,
  \href{http://dx.doi.org/10.1140/epjc/s10052-012-1965-6}{\doi{10.1140/epjc/s10052-012-1965-6}},
\href{http://www.arXiv.org/abs/1201.5595}{\texttt{arXiv:1201.5595}}.

\bibitem{Aad:2013gva}
\hrefCMSnoop {}{{ATLAS Collaboration}, ``{Search for long-lived stopped
  \textit{R}-hadrons decaying out-of-time with pp collisions using the ATLAS
  detector}'',} \textit{ Phys. Rev. D} \textbf{ 88} (2013) 112003,
  \href{http://dx.doi.org/10.1103/PhysRevD.88.112003}{\doi{10.1103/PhysRevD.88.112003}},
\href{http://www.arXiv.org/abs/1310.6584}{\texttt{arXiv:1310.6584}}.

\bibitem{Khachatryan:2015jha}
\hrefCMSnoop {}{{CMS Collaboration}, ``{Search for decays of stopped long-lived
  particles produced in proton-proton collisions at $\sqrt{s} = 8\TeV$}'',}
  \textit{ Eur. Phys. J. C} \textbf{ 75} (2015) 151,
  \href{http://dx.doi.org/10.1140/epjc/s10052-015-3367-z}{\doi{10.1140/epjc/s10052-015-3367-z}},
\href{http://www.arXiv.org/abs/1501.05603}{\texttt{arXiv:1501.05603}}.

\bibitem{Aad:2015rba}
\hrefCMSnoop {}{{ATLAS Collaboration}, ``{Search for massive, long-lived
  particles using multitrack displaced vertices or displaced lepton pairs in pp
  collisions at $\sqrt{s} = 8\TeV$ with the ATLAS detector}'',} \textit{ Phys.
  Rev. D} \textbf{ 92} (2015) 072004,
  \href{http://dx.doi.org/10.1103/PhysRevD.92.072004}{\doi{10.1103/PhysRevD.92.072004}},
\href{http://www.arXiv.org/abs/1504.05162}{\texttt{arXiv:1504.05162}}.

\bibitem{Aaboud:2016dgf}
\hrefCMSnoop {}{{ATLAS Collaboration}, ``{Search for metastable heavy charged
  particles with large ionization energy loss in pp collisions at $\sqrt{s} =
  13$ TeV using the ATLAS experiment}'',} \textit{ Phys. Rev. D} \textbf{ 93}
  (2016) 112015,
  \href{http://dx.doi.org/10.1103/PhysRevD.93.112015}{\doi{10.1103/PhysRevD.93.112015}},
\href{http://www.arXiv.org/abs/1604.04520}{\texttt{arXiv:1604.04520}}.

\bibitem{Aaboud:2017iio}
\hrefCMSnoop {}{{ATLAS Collaboration}, ``{Search for long-lived, massive
  particles in events with displaced vertices and missing transverse momentum
  in $\sqrt{s}$ = 13 TeV pp collisions with the ATLAS detector}'',} \textit{
  Phys. Rev. D} \textbf{ 97} (2018) 052012,
  \href{http://dx.doi.org/10.1103/PhysRevD.97.052012}{\doi{10.1103/PhysRevD.97.052012}},
\href{http://www.arXiv.org/abs/1710.04901}{\texttt{arXiv:1710.04901}}.

\bibitem{Sirunyan2018}
\hrefCMSnoop {}{{CMS Collaboration}, ``{Search for decays of stopped exotic
  long-lived particles produced in proton-proton collisions at $\sqrt{s}=$ 13
  TeV}'',} \textit{ JHEP} \textbf{ 05} (2018) 127,
  \href{http://dx.doi.org/10.1007/JHEP05(2018)127}{\doi{10.1007/JHEP05(2018)127}},
\href{http://www.arXiv.org/abs/1801.00359}{\texttt{arXiv:1801.00359}}.

\bibitem{Aaboud:2019opc}
\hrefCMSnoop {}{{ATLAS Collaboration}, ``{Search for long-lived neutral
  particles in $pp$ collisions at $\sqrt{s}$ = 13 TeV that decay into displaced
  hadronic jets in the ATLAS calorimeter}'',} \textit{ {Eur. Phys. J. C}}
  \textbf{ 79} (2019) 481,
  \href{http://dx.doi.org/10.1140/epjc/s10052-019-6962-6}{\doi{10.1140/epjc/s10052-019-6962-6}},
\href{http://www.arXiv.org/abs/1902.03094}{\texttt{arXiv:1902.03094}}.

\bibitem{displaced-dijets}
\hrefCMSnoop {}{{CMS Collaboration}, ``{Search for long-lived particles
  decaying into displaced jets in proton-proton collisions at $\sqrt{s}=$ 13
  TeV}'',} \textit{ Phys. Rev. D} \textbf{ 99} (2019) 032011,
  \href{http://dx.doi.org/10.1103/PhysRevD.99.032011}{\doi{10.1103/PhysRevD.99.032011}},
\href{http://www.arXiv.org/abs/1811.07991}{\texttt{arXiv:1811.07991}}.

\bibitem{Sirunyan:2018pwn}
\hrefCMSnoop {}{{CMS Collaboration}, ``{Search for long-lived particles with
  displaced vertices in multijet events in proton-proton collisions at
  $\sqrt{s} = 13$ TeV}'',} \textit{ Phys. Rev. D} \textbf{ 98} (2018) 092011,
  \href{http://dx.doi.org/10.1103/PhysRevD.98.092011}{\doi{10.1103/PhysRevD.98.092011}},
\href{http://www.arXiv.org/abs/1808.03078}{\texttt{arXiv:1808.03078}}.

\bibitem{Aaij2017}
\hrefCMSnoop {}{{LHCb Collaboration}, ``Updated search for long-lived particles
  decaying to jet pairs'',} \textit{ Eur. Phys. J. C} \textbf{ 77} (2017) 812,
  \href{http://dx.doi.org/10.1140/epjc/s10052-017-5178-x}{\doi{10.1140/epjc/s10052-017-5178-x}},
  \href{http://www.arXiv.org/abs/1705.07332}{\texttt{arXiv:1705.07332}}.

\bibitem{ATLAS:2012av}
\hrefCMSnoop {}{{ATLAS Collaboration}, ``{Search for a light Higgs boson
  decaying to long-lived weakly-interacting particles in proton-proton
  collisions at $\sqrt{s}=7$ TeV with the ATLAS detector}'',} \textit{ Phys.
  Rev. Lett.} \textbf{ 108} (2012) 251801,
  \href{http://dx.doi.org/10.1103/PhysRevLett.108.251801}{\doi{10.1103/PhysRevLett.108.251801}},
\href{http://www.arXiv.org/abs/1203.1303}{\texttt{arXiv:1203.1303}}.

\bibitem{Aad:2015uaa}
\hrefCMSnoop {}{{ATLAS Collaboration}, ``{Search for long-lived, weakly
  interacting particles that decay to displaced hadronic jets in proton-proton
  collisions at $\sqrt{s}=8$ TeV with the ATLAS detector}'',} \textit{ Phys.
  Rev. D} \textbf{ 92} (2015), no.~1, 012010,
  \href{http://dx.doi.org/10.1103/PhysRevD.92.012010}{\doi{10.1103/PhysRevD.92.012010}},
\href{http://www.arXiv.org/abs/1504.03634}{\texttt{arXiv:1504.03634}}.

\bibitem{Aaboud:2018aqj}
\hrefCMSnoop {}{{ATLAS Collaboration}, ``{Search for long-lived particles
  produced in $pp$ collisions at $\sqrt{s}=13$ TeV that decay into displaced
  hadronic jets in the ATLAS muon spectrometer}'',} \textit{ Phys. Rev. D}
  \textbf{ 99} (2019), no.~5, 052005,
  \href{http://dx.doi.org/10.1103/PhysRevD.99.052005}{\doi{10.1103/PhysRevD.99.052005}},
\href{http://www.arXiv.org/abs/1811.07370}{\texttt{arXiv:1811.07370}}.

\bibitem{CMS-displaced-photons}
\hrefCMSnoop {}{{CMS Collaboration}, ``{Search for long-lived particles in
  events with photons and missing energy in proton-proton collisions at
  $\sqrt{s}=7$ \TeV}'',} \textit{ Phys. Lett. B} \textbf{ 722} (2013) 273,
  \href{http://dx.doi.org/10.1016/j.physletb.2013.04.027}{\doi{10.1016/j.physletb.2013.04.027}},
  \href{http://www.arXiv.org/abs/1212.1838}{\texttt{arXiv:1212.1838}}.

\bibitem{ATLAS-displaced-photons}
\hrefCMSnoop {}{{ATLAS Collaboration}, ``{Search for nonpointing and delayed
  photons in the diphoton and missing transverse momentum final state in 8 TeV
  pp collisions at the LHC using the ATLAS detector}'',} \textit{ Phys. Rev. D}
  \textbf{ 90} (2014) 112005,
  \href{http://dx.doi.org/10.1103/PhysRevD.90.112005}{\doi{10.1103/PhysRevD.90.112005}},
\href{http://www.arXiv.org/abs/1409.5542}{\texttt{arXiv:1409.5542}}.

\bibitem{Liu:2018wte}
\hrefCMSnoop {}{J.~Liu, Z.~Liu, and L.-T. Wang, ``{Long-lived particles at the
  LHC: catching them in time}'',} \textit{ Phys. Rev. Lett.} \textbf{ 122}
  (2019) 131801,
  \href{http://dx.doi.org/10.1103/PhysRevLett.122.131801}{\doi{10.1103/PhysRevLett.122.131801}},
\href{http://www.arXiv.org/abs/1805.05957}{\texttt{arXiv:1805.05957}}.

\bibitem{Chatrchyan:2014fea}
\hrefCMSnoop {}{{CMS Collaboration}, ``{Description and performance of track
  and primary-vertex reconstruction with the CMS tracker}'',} \textit{ JINST}
  \textbf{ 9} (2014) P10009,
  \href{http://dx.doi.org/10.1088/1748-0221/9/10/P10009}{\doi{10.1088/1748-0221/9/10/P10009}},
\href{http://www.arXiv.org/abs/1405.6569}{\texttt{arXiv:1405.6569}}.

\bibitem{Chatrchyan:2009ih}
\hrefCMSnoop {}{{CMS Collaboration}, ``Calibration of the {CMS} drift tube
  chambers and measurement of the drift velocity with cosmic rays'',} \textit{
  JINST} \textbf{ 5} (2010) T03016,
  \href{http://dx.doi.org/10.1088/1748-0221/5/03/T03016}{\doi{10.1088/1748-0221/5/03/T03016}},
\href{http://www.arXiv.org/abs/0911.4895}{\texttt{arXiv:0911.4895}}.

\bibitem{timing-note}
\hrefCMSnoop {}{D.~del Re, ``{Timing performance of the CMS ECAL and prospects
  for the future}'',} \textit{ Journal of Physics: Conference Series} \textbf{
  587} (2015) 012003,
  \href{http://dx.doi.org/10.1088/1742-6596/587/1/012003}{\doi{10.1088/1742-6596/587/1/012003}}.

\bibitem{Khachatryan:2016bia}
\hrefCMSnoop {}{{CMS Collaboration}, ``{The CMS trigger system}'',} \textit{
  JINST} \textbf{ 12} (2017) P01020,
  \href{http://dx.doi.org/10.1088/1748-0221/12/01/P01020}{\doi{10.1088/1748-0221/12/01/P01020}},
\href{http://www.arXiv.org/abs/1609.02366}{\texttt{arXiv:1609.02366}}.

\bibitem{Cacciari:2008gp}
\hrefCMSnoop {}{M.~Cacciari, G.~P. Salam, and G.~Soyez, ``{The anti-\kt jet
  clustering algorithm}'',} \textit{ JHEP} \textbf{ 04} (2008) 063,
  \href{http://dx.doi.org/10.1088/1126-6708/2008/04/063}{\doi{10.1088/1126-6708/2008/04/063}},
\href{http://www.arXiv.org/abs/0802.1189}{\texttt{arXiv:0802.1189}}.

\bibitem{Cacciari:2011ma}
\hrefCMSnoop {}{M.~Cacciari, G.~P. Salam, and G.~Soyez, ``{FastJet user
  manual}'',} \textit{ Eur. Phys. J. C} \textbf{ 72} (2012) 1896,
  \href{http://dx.doi.org/10.1140/epjc/s10052-012-1896-2}{\doi{10.1140/epjc/s10052-012-1896-2}},
\href{http://www.arXiv.org/abs/1111.6097}{\texttt{arXiv:1111.6097}}.

\bibitem{Khachatryan:2016kdb}
\hrefCMSnoop {}{{CMS Collaboration}, ``{Jet energy scale and resolution in the
  CMS experiment in pp collisions at 8 TeV}'',} \textit{ JINST} \textbf{ 12}
  (2017) P02014,
  \href{http://dx.doi.org/10.1088/1748-0221/12/02/P02014}{\doi{10.1088/1748-0221/12/02/P02014}},
\href{http://www.arXiv.org/abs/1607.03663}{\texttt{arXiv:1607.03663}}.

\bibitem{CMS-PRF-14-001}
\hrefCMSnoop {}{{CMS Collaboration}, ``Particle-flow reconstruction and global
  event description with the {CMS} detector'',} \textit{ JINST} \textbf{ 12}
  (2017) P10003,
  \href{http://dx.doi.org/10.1088/1748-0221/12/10/P10003}{\doi{10.1088/1748-0221/12/10/P10003}},
\href{http://www.arXiv.org/abs/1706.04965}{\texttt{arXiv:1706.04965}}.

\bibitem{CMS-PAS-LUM-17-001}
\href {https://cds.cern.ch/record/2257069}{{CMS Collaboration}, ``{CMS
  luminosity measurement for the 2016 data taking period}'',} CMS Physics
  Analysis Summary CMS-PAS-LUM-17-001, 2017.

\bibitem{CMS-PAS-LUM-17-004}
\href {https://cds.cern.ch/record/2621960}{{CMS Collaboration}, ``{CMS
  luminosity measurement for the 2017 data-taking period at $\sqrt{s} =
  13~\mathrm{TeV}$}'',} CMS Physics Analysis Summary CMS-PAS-LUM-17-004, 2018.

\bibitem{CMS-PAS-LUM-18-002}
\href {http://cds.cern.ch/record/2676164}{{CMS Collaboration}, ``{CMS
  luminosity measurement for the 2018 data-taking period at $\sqrt{s} =
  13~\mathrm{TeV}$}'',} CMS Physics Analysis Summary CMS-PAS-LUM-18-002, 2019.

\bibitem{Sirunyan:2019kia}
\hrefCMSnoop {}{{CMS Collaboration}, ``Performance of missing transverse
  momentum reconstruction in proton-proton collisions at $\sqrt{s} = 13$\,{TeV}
  using the {CMS} detector'',} \textit{ JINST} \textbf{ 14} (2019) P07004,
  \href{http://dx.doi.org/10.1088/1748-0221/14/07/P07004}{\doi{10.1088/1748-0221/14/07/P07004}},
\href{http://www.arXiv.org/abs/1903.06078}{\texttt{arXiv:1903.06078}}.

\bibitem{Beenakker:1996ch}
\hrefCMSnoop {}{W.~Beenakker, R.~Hopker, M.~Spira, and P.~M. Zerwas, ``{Squark
  and gluino production at hadron colliders}'',} \textit{ Nucl. Phys. B}
  \textbf{ 492} (1997) 51,
  \href{http://dx.doi.org/10.1016/S0550-3213(97)80027-2}{\doi{10.1016/S0550-3213(97)80027-2}},
\href{http://www.arXiv.org/abs/hep-ph/9610490}{\texttt{arXiv:hep-ph/9610490}}.

\bibitem{Kulesza:2008jb}
\hrefCMSnoop {}{A.~Kulesza and L.~Motyka, ``{Threshold resummation for
  squark-antisquark and gluino-pair production at the LHC}'',} \textit{ Phys.
  Rev. Lett.} \textbf{ 102} (2009) 111802,
  \href{http://dx.doi.org/10.1103/PhysRevLett.102.111802}{\doi{10.1103/PhysRevLett.102.111802}},
\href{http://www.arXiv.org/abs/0807.2405}{\texttt{arXiv:0807.2405}}.

\bibitem{Kulesza:2009kq}
\hrefCMSnoop {}{A.~Kulesza and L.~Motyka, ``{Soft gluon resummation for the
  production of gluino-gluino and squark-antisquark pairs at the LHC}'',}
  \textit{ Phys. Rev. D} \textbf{ 80} (2009) 095004,
  \href{http://dx.doi.org/10.1103/PhysRevD.80.095004}{\doi{10.1103/PhysRevD.80.095004}},
\href{http://www.arXiv.org/abs/0905.4749}{\texttt{arXiv:0905.4749}}.

\bibitem{Beenakker:2009ha}
W.~Beenakker\hrefCMSnoop {}{ {et~al.}, ``{Soft-gluon resummation for squark and
  gluino hadroproduction}'',} \textit{ JHEP} \textbf{ 12} (2009) 041,
  \href{http://dx.doi.org/10.1088/1126-6708/2009/12/041}{\doi{10.1088/1126-6708/2009/12/041}},
\href{http://www.arXiv.org/abs/0909.4418}{\texttt{arXiv:0909.4418}}.

\bibitem{Beenakker:2011fu}
W.~Beenakker\hrefCMSnoop {}{ {et~al.}, ``Squark and gluino hadroproduction'',}
  \textit{ Int. J. Mod. Phys. A} \textbf{ 26} (2011) 2637,
  \href{http://dx.doi.org/10.1142/S0217751X11053560}{\doi{10.1142/S0217751X11053560}},
\href{http://www.arXiv.org/abs/1105.1110}{\texttt{arXiv:1105.1110}}.

\bibitem{Borschensky:2014cia}
C.~Borschensky\hrefCMSnoop {}{ {et~al.}, ``{Squark and gluino production cross
  sections in pp collisions at $\sqrt{s}$ = 13, 14, 33 and 100 TeV}'',}
  \textit{ Eur. Phys. J. C} \textbf{ 74} (2014) 3174,
  \href{http://dx.doi.org/10.1140/epjc/s10052-014-3174-y}{\doi{10.1140/epjc/s10052-014-3174-y}},
\href{http://www.arXiv.org/abs/1407.5066}{\texttt{arXiv:1407.5066}}.

\bibitem{Beenakker2016}
W.~Beenakker\hrefCMSnoop {}{ {et~al.}, ``{NNLL-fast: predictions for coloured
  supersymmetric particle production at the LHC with threshold and Coulomb
  resummation}'',} \textit{ JHEP} \textbf{ 12} (2016) 133,
  \href{http://dx.doi.org/10.1007/JHEP12(2016)133}{\doi{10.1007/JHEP12(2016)133}},
  \href{http://www.arXiv.org/abs/1607.07741}{\texttt{arXiv:1607.07741}}.

\bibitem{pythia}
\hrefCMSnoop {}{T.~Sj{\"o}strand, S.~Mrenna, and P.~Z. Skands, ``{A brief
  introduction to PYTHIA 8.1}'',} \textit{ Comput. Phys. Commun.} \textbf{ 178}
  (2008) 852,
  \href{http://dx.doi.org/10.1016/j.cpc.2008.01.036}{\doi{10.1016/j.cpc.2008.01.036}},
\href{http://www.arXiv.org/abs/0710.3820}{\texttt{arXiv:0710.3820}}.

\bibitem{nnpdf}
\hrefCMSnoop {}{{NNPDF} Collaboration, ``Parton distributions for the {LHC Run
  II}'',} \textit{ JHEP} \textbf{ 04} (2015) 040,
  \href{http://dx.doi.org/10.1007/JHEP04(2015)040}{\doi{10.1007/JHEP04(2015)040}},
\href{http://www.arXiv.org/abs/1410.8849}{\texttt{arXiv:1410.8849}}.

\bibitem{Fairbairn:2006gg}
M.~Fairbairn\hrefCMSnoop {}{ {et~al.}, ``{Stable massive particles at
  colliders}'',} \textit{ Phys. Rept.} \textbf{ 438} (2007) 1,
  \href{http://dx.doi.org/10.1016/j.physrep.2006.10.002}{\doi{10.1016/j.physrep.2006.10.002}},
\href{http://www.arXiv.org/abs/hep-ph/0611040}{\texttt{arXiv:hep-ph/0611040}}.

\bibitem{Kraan:2004tz}
\hrefCMSnoop {}{A.~C. Kraan, ``{Interactions of heavy stable hadronizing
  particles}'',} \textit{ Eur. Phys. J. C} \textbf{ 37} (2004) 91,
  \href{http://dx.doi.org/10.1140/epjc/s2004-01946-6}{\doi{10.1140/epjc/s2004-01946-6}},
\href{http://www.arXiv.org/abs/hep-ex/0404001}{\texttt{arXiv:hep-ex/0404001}}.

\bibitem{rhadorig}
\hrefCMSnoop {}{G.~Farrar and P.~Fayet, ``Phenomenology of the production,
  decay, and detection of new hadronic states associated with supersymmetry'',}
  \textit{ Phys. Lett. B} \textbf{ 76} (1978) 575,
  \href{http://dx.doi.org/10.1016/0370-2693(78)90858-4}{\doi{10.1016/0370-2693(78)90858-4}}.

\bibitem{CP2}
\hrefCMSnoop {}{{CMS Collaboration}, ``{Extraction and validation of a new set
  of CMS PYTHIA8 tunes from underlying-event measurements}'',} (2019).
  \href{http://www.arXiv.org/abs/1903.12179}{\texttt{arXiv:1903.12179}}.
Submitted to {\it EPJC}.

\bibitem{Alwall2014}
J.~Alwall\hrefCMSnoop {}{ {et~al.}, ``The automated computation of tree-level
  and next-to-leading order differential cross sections, and their matching to
  parton shower simulations'',} \textit{ JHEP} \textbf{ 07} (2014) 079,
  \href{http://dx.doi.org/10.1007/JHEP07(2014)079}{\doi{10.1007/JHEP07(2014)079}},
  \href{http://www.arXiv.org/abs/1405.0301}{\texttt{arXiv:1405.0301}}.

\bibitem{geant}
\hrefCMSnoop {}{{GEANT4} Collaboration, ``{GEANT4}---a simulation toolkit'',}
  \textit{ Nucl. Instr. Meth. A} \textbf{ 506} (2003) 250,
\href{http://dx.doi.org/10.1016/S0168-9002(03)01368-8}{\doi{10.1016/S0168-9002(03)01368-8}}.

\bibitem{1590608}
A.~{Bertarelli}\hrefCMSnoop {}{ {et~al.}, ``Mechanical design for robustness of
  the {LHC} collimators'',} \textit{ Proceedings of the 2005 Particle
  Accelerator Conference} (2005) 913,
  \href{http://dx.doi.org/10.1109/PAC.2005.1590608}{\doi{10.1109/PAC.2005.1590608}}.

\bibitem{Mackeprang:2006gx}
\hrefCMSnoop {}{R.~Mackeprang and A.~Rizzi, ``Interactions of coloured heavy
  stable particles in matter'',} \textit{ Eur. Phys. J. C} \textbf{ 50} (2007)
  353,
  \href{http://dx.doi.org/10.1140/epjc/s10052-007-0252-4}{\doi{10.1140/epjc/s10052-007-0252-4}},
\href{http://www.arXiv.org/abs/hep-ph/0612161}{\texttt{arXiv:hep-ph/0612161}}.

\bibitem{inelast}
\hrefCMSnoop {}{{CMS Collaboration}, ``{Measurement of the inelastic
  proton-proton cross section at $ \sqrt{s}=13 $ TeV}'',} \textit{ JHEP}
  \textbf{ 07} (2018) 161,
  \href{http://dx.doi.org/10.1007/JHEP07(2018)161}{\doi{10.1007/JHEP07(2018)161}},
\href{http://www.arXiv.org/abs/1802.02613}{\texttt{arXiv:1802.02613}}.

\bibitem{CMS-NOTE-2011-005}
\href {https://cds.cern.ch/record/1379837}{{ATLAS and CMS Collaborations},
  ``{Procedure for the LHC Higgs boson search combination in Summer 2011}'',}
  Technical Report CMS-NOTE-2011-005, ATL-PHYS-PUB-2011-11, 2011.

\bibitem{junk}
\hrefCMSnoop {}{T.~Junk, ``Confidence level computation for combining searches
  with small statistics'',} \textit{ Nucl. Instr. Meth. A} \textbf{ 434} (1999)
  435,
  \href{http://dx.doi.org/10.1016/S0168-9002(99)00498-2}{\doi{10.1016/S0168-9002(99)00498-2}},
  \href{http://www.arXiv.org/abs/hep-ex/9902006}{\texttt{arXiv:hep-ex/9902006}}.

\bibitem{CLsTechnique}
\hrefCMSnoop {}{A.~L. Read, ``{Presentation of search results: the CLs
  technique}'',} \textit{ J. Phys. G} \textbf{ 28} (2002) 2693,
\href{http://dx.doi.org/10.1088/0954-3899/28/10/313}{\doi{10.1088/0954-3899/28/10/313}}.

\end{thebibliography}\endgroup

\cleardoublepage \appendix\section{The CMS Collaboration \label{app:collab}}\begin{sloppypar}\hyphenpenalty=5000\widowpenalty=500\clubpenalty=5000\vskip\cmsinstskip
\textbf{Yerevan Physics Institute, Yerevan, Armenia}\\*[0pt]
A.M.~Sirunyan$^{\textrm{\dag}}$, A.~Tumasyan
\vskip\cmsinstskip
\textbf{Institut für Hochenergiephysik, Wien, Austria}\\*[0pt]
W.~Adam, F.~Ambrogi, T.~Bergauer, J.~Brandstetter, M.~Dragicevic, J.~Erö, A.~Escalante~Del~Valle, M.~Flechl, R.~Frühwirth\cmsAuthorMark{1}, M.~Jeitler\cmsAuthorMark{1}, N.~Krammer, I.~Krätschmer, D.~Liko, T.~Madlener, I.~Mikulec, N.~Rad, J.~Schieck\cmsAuthorMark{1}, R.~Schöfbeck, M.~Spanring, D.~Spitzbart, W.~Waltenberger, C.-E.~Wulz\cmsAuthorMark{1}, M.~Zarucki
% ...
\vskip\cmsinstskip
\dag: Deceased\\
1:  Also at Vienna University of Technology, Vienna, Austria\\
% ...
\end{sloppypar}
\end{document}